\documentclass[sigplan,screen]{acmart}

\renewcommand\footnotetextcopyrightpermission[1]{} % removes footnote with conference information in first column
\setcopyright{none}

\usepackage{times}  % DO NOT CHANGE THIS
\usepackage{helvet}  % DO NOT CHANGE THIS
\usepackage{courier}  % DO NOT CHANGE THIS
\usepackage{graphicx} % DO NOT CHANGE THIS
\usepackage{caption} % DO NOT CHANGE THIS AND DO NOT ADD ANY OPTIONS TO IT
\DeclareCaptionStyle{ruled}{labelfont=normalfont,labelsep=colon,strut=off} % DO NOT CHANGE THIS
\usepackage{algorithm}
\usepackage{algorithmic}
\usepackage{multicol}
\usepackage{tikz}
\usepackage{amsmath,amsfonts}
\usepackage{mathtools}
\usepackage[utf8]{inputenc}
\usepackage[T1]{fontenc}
\usepackage{booktabs}
\usepackage{multirow} 

\usepackage{etoolbox,refcount}

\newcounter{countitems}
\newcounter{nextitemizecount}
\newcommand{\setupcountitems}{%
  \stepcounter{nextitemizecount}%
  \setcounter{countitems}{0}%
  \preto\item{\stepcounter{countitems}}%
}
\makeatletter
\newcommand{\computecountitems}{%
  \edef\@currentlabel{\number\c@countitems}%
  \label{countitems@\number\numexpr\value{nextitemizecount}-1\relax}%
}
\newcommand{\nextitemizecount}{%
  \getrefnumber{countitems@\number\c@nextitemizecount}%
}
\newcommand{\previtemizecount}{%
  \getrefnumber{countitems@\number\numexpr\value{nextitemizecount}-1\relax}%
}
\makeatother    
{\end{itemize}%
\unskip\computecountitems\ifnumcomp{\previtemizecount}{>}{3}{\end{multicols}}{}}

\usepackage{newfloat}
\usepackage{listings}
\floatstyle{ruled}
\newfloat{listing}{tb}{lst}{}
\floatname{listing}{Listing}

\title{"I'm Not for Sale" -- Perceptions and Limited Awareness of Privacy Risks by Digital Natives About Location Data}

\author{Antoine Boutet}
\affiliation{
	\institution{Univ Lyon, INSA Lyon, Inria, CITI}
    \country{} 
}
\email{antoine.boutet@insa-lyon.fr}

\author{Victor Morel}
\affiliation{%
  \institution{Chalmers University of Technology and University of Gothenburg}
  \country{} 
  }
\email{morelv@chalmers.se}

\usepackage{bibentry}
\begin{document}

\begin{abstract}
Although mobile devices benefit users in their daily lives in numerous ways, they also raise several privacy concerns.
For instance, they can reveal sensitive information that can be inferred from location data. 
This location data is shared through service providers as well as mobile applications.
Understanding how and with whom users share their location data as well as users' perception of the underlying privacy risks, are important notions to grasp in order to design usable privacy-enhancing technologies.
% likely to be adopted. 
In this work, we perform a quantitative and qualitative analysis of smartphone users’ awareness, perception and self-reported behavior towards location data-sharing through a survey of $n=99$ young adult participants (i.e., digital natives).
We compare stated practices with actual behaviors to better understand their mental models, and survey participants' understanding of privacy risks before and after the inspection of location traces and the information that can be inferred therefrom.

Our empirical results show that participants have risky privacy practices: about 54\% of participants underestimate the number of mobile applications to which they have granted access to their data, and 33\% forget or do not think of revoking access to their data.
Furthermore, most of the participants do not have a realistic perception of privacy risks and have generally heard little about privacy-related scandals.
% , apart from the case of Cambridge Analytica. 
Also, by using a demonstrator to perform inferences from location data, we observe that slightly more than half of participants (57\%) are surprised by the extent of potentially inferred information, and that 47\% intend to reduce access to their data via permissions as a result of using the demonstrator.
Last, a majority of participants have little knowledge of the tools to better protect themselves, but are nonetheless willing to follow suggestions to improve privacy (51\%).
Educating people, including digital natives, about privacy risks through transparency tools seems a promising approach.
\end{abstract}

\maketitle
\pagestyle{plain}

\section{Introduction}
\label{sec:intro}

Smartphones have become the most popular electronic devices, used by 95\% of all internet users\footnote{\url{https://datareportal.com/global-digital-overview}}.
The widespread adoption of these mobile devices with geolocation capabilities
% (what we denote as \textit{location contexts} in this work) 
makes users' location traces\footnote{That is, location data collected by Google from smartphones which gives more accurate and reliable information than other kinds of location data collection} widely accessible to mobile services.
Location is the most collected personal data by mobile applications~\cite{achara:hal-00917417}, and while the exploitation of this data presents an obvious benefit to users in their daily lives -- through personalized services, and potentially to the society with crowdsourcing initiatives~\cite{stevens2010crowdsourcing} --, an uncontrolled exploitation of this data continue to raise privacy concerns.

%https://www.theguardian.com/technology/2014/jun/27/new-york-taxi-details-anonymised-data-researchers-warn

This location data can indeed reveal a lot of sensitive information about users and represents unique privacy risks and implications compared to other types of data shared via smartphones.
%This data can indeed reveal a lot of sensitive information about users. 
For example, it has been shown that it is possible to infer personality traits, religious affiliation, sexual orientation, or health status~\cite{10.1145/1868470.1868479}.
Additionally, the uniqueness of location data provides a sort of fingerprint specific to each individual that can be used to re-identify users in anonymized data, i.e., little \textit{a priori} knowledge about a user can be exploited to discriminate them in a large set of anonymous location data.
For instance, journalists were able to re-identify and track the whereabouts of the former US president Trump from a large dataset~\cite{trump}, and anonymous data was used to pinpoint Muslim cab drivers~\cite{cab}.
The attack surface of the exploitation of personal data is not fully discovered.  
Regularly, new threats appear and are maliciously exploited for influential campaigns and interfering in political election\footnote{\url{https://en.wikipedia.org/wiki/Troll_farm}}, or discovering sensitive information\footnote{\url{https://www.nytimes.com/2018/01/29/world/middleeast/strava-heat-map.html}}.
%.the position of military bases\footnote{\url{https://www.nytimes.com/2018/01/29/world/middleeast/strava-heat-map.html}}.

With this increasing exposure of privacy risks, understanding how young users share their data, and the extent to which they are aware of security and privacy risks, are important properties to assess these risks and to develop effective Privacy and Transparency Enhancing Technologies %(PETs and TETs, respectively)
matching current users' expectations to increase adoption.
Although user perceptions of technology~\cite{1203752} and the privacy paradox\footnote{The privacy paradox refers to self-reported concerns about privacy appear to be in contradiction with often careless online behaviors.}~\cite{BARTH201955, kang2021smart} have received a lot of attention, users’ self-reported behaviors in mobility contexts associated with smartphones, and the impact of an awareness demonstration platform %of TETs such as of demonstrators 
have been less studied in the academic literature.
%In order t
To fill this gap, this work addresses two research questions:
\begin{itemize}
    \item RQ1: What are the perceptions and the understanding of young users' privacy and its protection in a mobility context?
    \item RQ2: What is the impact of a demonstrator %TET 
    for the visualization of location traces and associated privacy risks on these perceptions and understanding?
\end{itemize}

In order to answer our research questions, we designed and deployed survey questionnaires answered by $n=99$ young participants (i.e., digital natives, persons who grew up in the information age, aged between 20 and 26 with an average of 21 in our case).
%we restricted the term ``users'' as a generic term for individuals).
% not necessarily participating in our study).
Specifically, we explored the participants' behavior, self-reported behavior, and awareness regarding their own data-sharing practices. 
We devised a first questionnaire to study their perception of privacy and to their permission management with respect to location, in which we emphasized the eventual discrepancies between their remembered practices and their actual behaviors.
% through several questions related to their 
% what they think they do, versus what they actually do.
We also surveyed participants' understanding of privacy risks before and after exposing them to location traces demonstrating what information can be inferred from this data, as well as their awareness of protection tools, in conjunction with a second questionnaire.
%We also surveyed participants' understanding of %the 
%privacy risks, before and after the inspection of location traces and the information that can be inferred therefrom, and their knowledge of protection tools, jointly %in combination 
%with a second questionnaire.

Our results show that participants have \textbf{risky practices in terms of privacy} where more than half of participants underestimate the number of mobile applications to which they have granted access to their data.
In addition, most of the participants tend to forget about disabling location access permission for apps that are not actively used.
Our results also show that participants are \textbf{poorly aware of the privacy risks} and are unable to list cases of personal data leaks or scandals linked to their uses despite the media coverage. 
Moreover, by using a demonstrator to perform inferences from location data, % (i.e., their own location traces or examples), 
more than half of the participants (57\%) are surprised by the extent of potentially inferred information and 47\% intend to reduce access to their data via permissions.
Finally, most of the participants are inclined to \textbf{better use protection tools in the future}, even if they are still little aware of the available tools to improve privacy today.
% \textcolor{red}{Since our study focused mainly on digital natives, the transferability of the results to the rest of the population remains an open question. However, the academic knowledge in computer science of our participants certainly gives them a better perception of the risk to privacy than among other age groups.}
Since our study focused mainly on digital natives, the transferability of the results to the rest of the population remains an open question. However, the academic knowledge in computer science of our participants certainly gives them a better perception of the risk to privacy than among other age groups.
Our contributions are the following:
\begin{enumerate}
    \item We provide the first insights on privacy perceptions in a mobility context associated with smartphones and mobile applications; % (to the best of our knowledge);
    \item We study the impact of a demonstration platform of privacy risks visualization %;TET 
    on privacy perceptions; % via a demonstration platform of privacy risks visualization;
    \item We formulate a set of recommendations to improve the management of privacy permissions on mobile.  
\end{enumerate}

The paper is organized as follows. 
Section~\ref{sec:related} reviews the related work.
Section~\ref{sec:methodology} describes the considered methodology while the results are presented in Section~\ref{sec:results}.
We discuss recommendations and highlight our limitations in Section~\ref{sec:discussion}, and conclude in Section~\ref{sec:conclusion}.

\section{Related Work}
\label{sec:related}

%\todo[inline]{@Antoine: Streamline related work. Add references connected to IUIPC, in WWW proceedings, about our population, and other relatable refs.}
% \todo[inline]{Antoine: privacy in location contexts/IoT}
%\todo[inline]{Antoine: double-check consistency and organization of the section please}

Prior research widely 
% \textcolor{red}{widely} 
explored location privacy (%presented in 
Section~\ref{sec:relatedPrivacy}) and users' perceptions of privacy (%presented in 
Section~\ref{sec:relatedPerception}).
% \textcolor{red}{However, little attention has been devoted to self-reported behaviors towards location data sharing. Moreover, to the best of our knowledge, no studies have used a demonstration platform to improve risk understanding in a way that would change users’ perceptions about the balance between the benefits of invasive technologies and potential risks.}
However, little attention has been devoted to self-reported behaviors towards location data sharing. Moreover, to the best of our knowledge, no studies have used a demonstration platform to improve risk understanding in a way that would change users’ perceptions about the balance between the benefits of invasive technologies and potential risks.

\subsection{Privacy and location}
\label{sec:relatedPrivacy}

The privacy issues raised by location data gained a lot of traction in the last decade~\cite{8482357}. 
In particular, user location traces extracted from various data have been shown to be highly unique~\cite{de2013unique,Zang:2011:ALD:2030613.2030630}. %10.1145/3465481.3470474,
This high uniqueness may act as a digital fingerprint and lead to the re-identification of users if their traces are associated with external knowledge.
This uniqueness does not only concern location traces but characterizes all traces generated by a human ~\cite{journals/popets/Kurtz16, eckersley2010unique, journals/popets/Overdorf16}, and can generate a risk of re-identification.
%% This uniqueness does not only concern location traces but characterizes all traces generated by a human ranging from the personalized settings of mobile devices~\cite{journals/popets/Kurtz16} or Web browsers~\cite{eckersley2010unique}, to the logs of in-car sensors~\cite{journals/popets/EnevTKK16}, phone calls~\cite{Zang:2011:ALD:2030613.2030630}, or the writing style of users on the Web~\cite{journals/popets/Overdorf16}.
Several cases of re-identification have been documented, for instance the re-identification of individuals from web search queries~\cite{article2},  %movie ratings~\cite{Narayanan:2008:RDL:1397759.1398064}, 
taxi logs~\cite{zhang2016inferring}, or the notorious case of Governor William Weld using medical information~\cite{article}.

%\subsection{Re-identification}
The uniqueness and risk of re-identification is not the unique threat related to location traces.
Recent works have demonstrated that location is a very rich contextual information and leads to a strong inferential potential in terms of information that can be learned about individuals~\cite{boutet:hal-02421828}. 
For instance, location traces can reveal Points Of Interest (POI) of users such as their home and workplaces~\cite{10.1145/1868470.1868479}, their race and gender~\cite{10.1145/2684822.2685287}, their social network~\cite{10.1145/2665943.2665960}, it can be used to predict their location patterns~\cite{sadilek2012far}, to link accounts of the same user across different datasets~\cite{10.1145/2872427.2883002}, and infer even more sensitive information such as their religion or personality traits\footnote{Note that the processing of such sensitive information is prohibited by Article 9(1) of the General Data Protection Regulation (GDPR) on the processing of special categories of personal data.}.

Due to the large adoption of mobile devices, the location data of users is extensively collected and shared~\cite{almuhimedi2015your,10.1145/2976749.2978313}.
% with and without the consent of individuals~\cite{10.1145/2976749.2978313}.
The uncontrolled usage of this information can have an important impact on users such as unfair price discrimination~\cite{10.1145/2390231.2390245}.
The increasing use of connected devices collecting personal data and the lack of transparency on how the data is actually exploited raise privacy concerns.
Only a handful of tools have been proposed to improve user awareness about the potential risk of revealing their location.
For instance,~\cite{Riederer2016FindYouAP} propose a tool to inspect the potential of location data, while~\cite{boutet:hal-02421828} show the impact of protection mechanisms of the inference capabilities using a demonstration platform.

%\subsection{Perceptions of location privacy and user studies}
%\subsection{Perceptions of privacy and user studies}
\subsection{Perceptions of privacy and privacy controls}
\label{sec:relatedPerception}

Giving users the benefits of location services on their mobile devices while preserving their privacy is an ongoing challenge, as evidenced by mobile OSs iterating on the user interfaces for location notices and control every few years.
For example, iOS allows users to select whether apps get fine- or coarse-grained location data\footnote{https://support.apple.com/guide/iphone/control-the-location-information-you-share-iph3dd5f9be/ios}, and will
present pop-ups when apps continually access location services in the background.
Similarly, Android apps now have to request background location access separately from general location usage\footnote{https://developer.android.com/about/versions/11/privacy/location},
and the OS
%\footnote{https://developer.android.com/about/versions/11/privacy/permissions\#auto-reset} 
 will automatically revoke unused permissions from apps\footnote{https://developer.android.com/about/versions/11/privacy/permissions}. 
Even though mobile operating systems regularly improve user interfaces, the opaque privacy controls of location services still face criticism\footnote{https://www.attorneygeneral.gov/taking-action/attorney-general-josh-shapiro-announces-391-million-settlement-with-google-over-location-tracking-practices/}.
% Balash et al.~
\cite{277130} explore the users' perceptions regarding access to Google accounts by mobile applications and formulate design recommendations to improve the current third-party management tools offered by Google, such as tracking recent access, automatically revoking access due to app disuse, and providing permission controls.
% Wijesekera et al.~
\cite{190982} analyze authorization preferences in different usage contexts and suggest determining the situations in which users would like to be confronted with security decisions.

Users' perceptions of the risks and benefits of technologies can determine their willingness to adopt them~\cite{EuroUSEC16}. 
More specifically, people are more likely to accept potentially invasive technology if they think its benefits will outweigh its potential risks~\cite{1203752}.
Due to the massive adoption of location-enabled mobile applications, this fact suggests that users' perception of this trade-off is more in favor of the benefits than the potential risks.
This attitude however depends on the perception of the said privacy risks.
Studies have shown that users of mobile phones are often unaware of the data collected by apps running on their devices, and that a majority of users restrict some of their permissions~\cite{almuhimedi2015your} following a better awareness of data collection.
Other studies explored the privacy paradox~\cite{BARTH201955, kang2021smart} where self-reported concerns about privacy appear to be in contradiction with often careless online behaviors.
However, another study focuses on university community~\cite{gamarrapercepcion} and show that this population of users does not have a genuine concern regarding the privacy of their geolocation data.
Note that this paper did not study location data in mobility contexts associated with smartphones as we do.

%measures the perception of location privacy of users with mobile devices and

%\todo[inline]{Next paragraph is out of place}
%The current paper studies the actual users' behavior about data-sharing of location traces.
%While our results comfort the privacy paradox widely observed~\cite{BARTH201955, kang2021smart} -- where self-reported concerns about privacy appear to be in contradiction with often careless online behaviors --, we also show that users tend to underestimate the risks, and with a better understanding about them, they want to reduce the data-sharing.

%\todo[inline]{Next paragraph appears disconnected from the rest}
% \todo[inline]{Victor: user studies about privacy perceptions}
%Users studies about privacy perceptions recently gained a lot of traction in the literature.

Analyzing behavior and understanding users' perceptions are also important notions to grasp in order to further design Privacy and Transparency Enhancing Technologies (PETs and TETs).
For instance, 
% Kaushik et al.~
\cite{274429} conducted an online survey to understand people’s perspectives on solely automated decision-making. They then formulate recommendations on how to design such systems.
In a similar line of work, 
% Islami et al.~
\cite{islami_capturing_2022} conducted in-depth semi-structured interviews with 17 Swedish drivers to analyse their privacy perceptions and preferences for intelligent transportation systems, then to provide recommendations for suitable predefined privacy options.
%For instance, \cite{10.1145/3366423.3380273} compared intention and perception in online discussion and showed that reducing misperception is an important factor to promote healthler conversations.
Several other works~\cite{DBLP:journals/popets/ZuffereyNHH23,DBLP:journals/imwut/VelykoivanenkoN21, 10.1145/3491102.3502136} studied the perceptions of privacy and utility of users related to fitness-trackers and demonstrated a high potential for data minimization (i.e., reducing the volume of data sent to service provider).
% Debatin et al.~
\cite{10.1111/j.1083-6101.2009.01494.x}, in turn, investigated Facebook users' awareness of privacy issues and perceived benefits and risks of utilizing Facebook and recommended better privacy protection, higher transparency and more education about the risks of posting personal information to reduce risky behavior. 
% Bielova et al.~
\cite{bielova:hal-04235032} analysed the behaviors of websites' visitors through a study of the impact of dark patterns on consent decisions. 
% and provided recommendations.
% Veys et al.~
\cite{274435} explored whether current data downloads (such as Google Takeout) actually achieve the transparency goals embodied by the right of access. Most participants indicated that current offerings need improvement to be useful, emphasizing the need for better filtration, visualization, and summarizing to help them hone in on key information.
However, none of these related work specifically targets users' behavior and perception about data-sharing of location data.
Although 
% Martin and Nissenbaum~
\cite{martin_what_2019} clearly addressed users' perceptions of location data, it is to be noted that it does so from an information science standpoint, %-- that is, 
closer to sociology and not from a UX/usability one.
It nonetheless offers a relevant account for the interested reader.

Related to location data, 
% Farke et al.~
\cite{274582} specifically analysed the user perceptions and reactions to Google's My Activity and how this web history dashboard increases or decreases end-users' concerns and benefits regarding data collection. Their results show that participants were surprised by the volume and detail of the collected data, but most of them were significantly more likely to be both 1) less concerned about data collection and 2) to view data collection more beneficially. However, this dashboard does not present any risks such as possible sensitive inferences associated with location traces or places visited. By also presenting the risks, our study shows that users are more concerned about privacy issues.

\section{Methodology}
\label{sec:methodology}
%As mentioned in Section~\ref{sec:intro}, 
We conducted two questionnaires on $n=99$ young participants aged between 20 and 26 (with an average around 21) to collect quantitative and qualitative data about the participants' perception and self-reported behavior regarding the sharing of location data.
%To collect quantitative and qualitative data about the users' behavior regarding the sharing of location data, we conducted two questionnaires on $n=99$ users about their privacy perceptions in a location context, i.e., about their Google Maps traces.

In a nutshell, the first questionnaire (hereinafter ``\textbf{behavior questionnaire}'') aimed at assessing their self-reported practices, perceptions, and pre-conceived ideas about privacy.
Following this first questionnaire, participants were then invited to use a demonstrator (denoted ``\textbf{demonstration}'' in what follows) which analyses location traces and reports several inferences as well as the impact of a protection mechanism using differential privacy (i.e., geo-indistiguishability~\cite{Andr_s_2013}).
The second questionnaire (hereinafter ``\textbf{perception questionnaire}'') addressed their assessment of the risks.

%In this section, we describe the recruitment of participants who answered the survey (Section~\ref{subsec:recruitement}), ethical considerations (Section~\ref{sec:ethics}), the design of the questionnaires (Section~\ref{subsec:questionnaires}), the demonstration platform we used to make participants aware of the privacy risks (Section~\ref{subsec:analysis}), the considered methodology to perform the thematic analysis of free text answered in the study (Section~\ref{sec:analysis}), and general statistics about our participants (Section~\ref{sec:general}).

%, and the procedure (Section~\ref{subsec:procedure}).
%\todo[inline]{What do you mean by procedure? Is it what I coined the perception questionnaire?}

\subsection{Recruitment}
\label{subsec:recruitement}
We recruited young participants for our survey via a course offered to engineering students at [anonymized].
This process of recruitment ensured a high number of participants,\footnote{(We address the limitations of our recruitment in Section~\ref{sec:limitations}).} all of them using their smartphones.
The participants were not compensated.
The participants had followed computer science courses including data science, AI, and introduction to the GDPR. 
By mainly recruiting young computer engineers with knowledge of technologies, we performed our study on a homogeneous population which consumes digital information quickly, and which are supposedly aware of privacy risks, hence providing an ``upper bound'' of privacy perceptions in our context.
%EESN on 4th year engineer student at Insa-Lyon, France.
% This course is a program on enjeux environemental and responsabilité societale and concerne all the students.

The study was performed over a four hours slot, during which students were successively invited to answer the behavior questionnaire, analyse privacy risks through the demonstration, and %finally 
answer the perception questionnaire.
%Note that, a
Although the teacher of the course is one of the authors of the current paper, participants were presented with a consent form at the beginning of the survey, and they were able to decline without penalty (specified orally);
% \textcolor{red}{(specified orally)}; 
only the experimentation of the demonstration platform was mandatory as part of their curriculum.
% \textcolor{red}{The participation was anonymous, voluntary, and the course did not include any grade (one participant actually refused to answer the questionnaire, without penalty).}
The participation was anonymous, voluntary, and the course did not include any grade (one participant actually refused to answer the questionnaire, without penalty).
% \todo[inline]{Justify why you recruited your participants from this pool. Perhaps because it was the easiest way to have a lot of participants. But of course not a good reason for reviewers. Try to justify why such a population you deemed to be suitable for your study and why recruiting like this}

% \todo[inline]{Was participation mandatory?}
% The remaining time has been dedicated to an analysis on the utility and privacy tradeoff through three teams split to conduct naive anonymisation of a dataset of location traces, a utility assessment and a privacy assessment.

\subsection{Ethical considerations}
\label{sec:ethics}
% In order to provide a balanced perspective, authors are required to include a statement about the potential broader impact of their work and ethical considerations. Papers without such a statement will be desk-rejected. This statement needs to be presented in a clearly marked paragraph/subsection/section in your paper. Authors should take care to discuss both positive and negative outcomes. Authors are also expected to describe steps taken to prevent or mitigate potential negative outcomes. For full papers that have collected new datasets and for dataset papers, discuss ethical considerations in the data collection process, considerations around release of the data, and both potentially positive and negative outcomes of its use by others.
The survey raises ethical issues in two main aspects.% expressed here.

First, the demonstration platform uses personal data, specifically location traces. 
This data is not necessarily sensitive in itself, but can act as a proxy for sensitive data. 
To reduce risks, participants 
% were informed through a privacy policy (developed with experts in usable privacy) of the research purpose of processing, and 
used traces from data of the authors of this study.
% [anonymized].
%
%First, the demonstration platform uses personal data, specifically location traces.
%This data is not necessarily sensitive in itself, but can act as a proxy for sensitive data.
%Participants were informed through a privacy policy (developed with experts in usable privacy) of the research purpose of processing, and they had the possibility not to use their own data to experiment on the platform.
%To avoid loading the server and wasting time calculating Points of Interests (POIs), collecting metadata, and calculating inferences, we encouraged the students to use the examples present on the platform. 
%Therefore, only a small number analyzed their own location traces.
The data used during the demonstration phase was encrypted server-side and was deleted once all data had been processed (calculating Points of Interests, collecting metadata, and calculating inferences).
% for the visualization tool. 
The DPO of the authors' institution has validated the demonstration platform.
% \todo[inline]{To be rephrased and made more specific}
%\todo[inline]{Ethics issue limited because no personal data from students, as mentioned in the limitations. Do we even need to specify anything if the platform didn't collect their data??}

Second, we elicited answers from participants, which is also personal data.
% As the academic institution in which the survey was performed does not have its own IRB,\footnote{The demonstration platform was devised at an earlier point in time and within an independent institution from the academic one hosting the survey.} w
We took as many precautions as needed to guarantee 1) informed consent (see Section~\ref{app:study} for the consent form, and as noted in Section~\ref{subsec:recruitement} they had the possibility to decline without any penalty), and 2) security of data storage and processing (%the 
authors only communicated using end-to-end encrypted tools, and the survey was conducted on an EU-based online tool stamped GDPR-compliant).
One participant refused to answer the questionnaires.

Both the demonstration platform (regarding its compliance with the GDPR and other data security regulations)
% \textcolor{red}{(regarding its compliance with the GDPR and other data security regulations)} 
and the questionnaires were approved by the IRB of the academic institution both hosting the survey and where the students were based.

% We believe the present work can foster positive outcomes related to usability of location-based systems by encouraging a more privacy-friendly design of such systems.
% As future work, we would like to study the data sharing behavior of users on other types of information present on the main platforms such as health information, or web searches, as well as their reaction to a personalized risk assessment.
% Indeed, highlighting personalized risk has a much greater impact on user awareness.
% For instance, in a previous version of the demo platform which included photos of POIs (this service is no more freely available from location services APIs), we observed a significant reaction from users who visualized their house or their various places of visit.

\subsection{Design of the Questionnaires}
\label{subsec:questionnaires}
% \todo[inline]{Summary of the study questionnaire's sections}
We shortly describe in this section the organization of the two questionnaires.
The interested reader will find their full content in Appendix~\ref{app:questionnaires}.
% It was designed to take around Z minutes. 
Each section is succinctly described in what follows:
\begin{itemize}
    \item \textbf{Introduction:} This section presents the study and asks the participants for their consent to participate in the study and to collect their answers for research purposes.
    \item \textbf{Data-Sharing:} This section inquires about the participants' attitudes about data sharing (e.g., which apps they think could access their location data, with whom, etc). 
    \item \textbf{Inferences:} This section questions their pre-conceived ideas about inferences from location data (e.g., \textit{from} which data could their location be inferred, or whether the location is sufficient to single them out in a dataset).
    \item \textbf{Authorisation and Control:} This section touches upon authorisation and control of permissions in a mobile context. Questions in this section are about their revoking or giving access permissions to apps, or their evaluation of the difficulty to effectuate this permission management.
    \item \textbf{Privacy Concerns:} This section simply performs the analysis of an Internet User’s Information Privacy Concerns (IUIPC) on a 7 Likert scale.
    \item \textbf{Demographics:} This %last 
    section elicits demographics data.
    %\item and a sixth optional section asked a drawing of their \textbf{mental models} of location data transfer. 
\end{itemize}

The perception questionnaire is much shorter and is composed of questions regarding the possible (malicious) inferences conducted on their location traces, the protection against these privacy violations, and their new expectations and suggestions following the experimentation with the demonstration platform.

\subsection{Risks analysis - demonstration}
\label{subsec:analysis}
% \todo[inline]{V: add description of the demo, including defense mechanisms}

\begin{figure}[t]
    \centering
    \includegraphics[width=.44\textwidth]{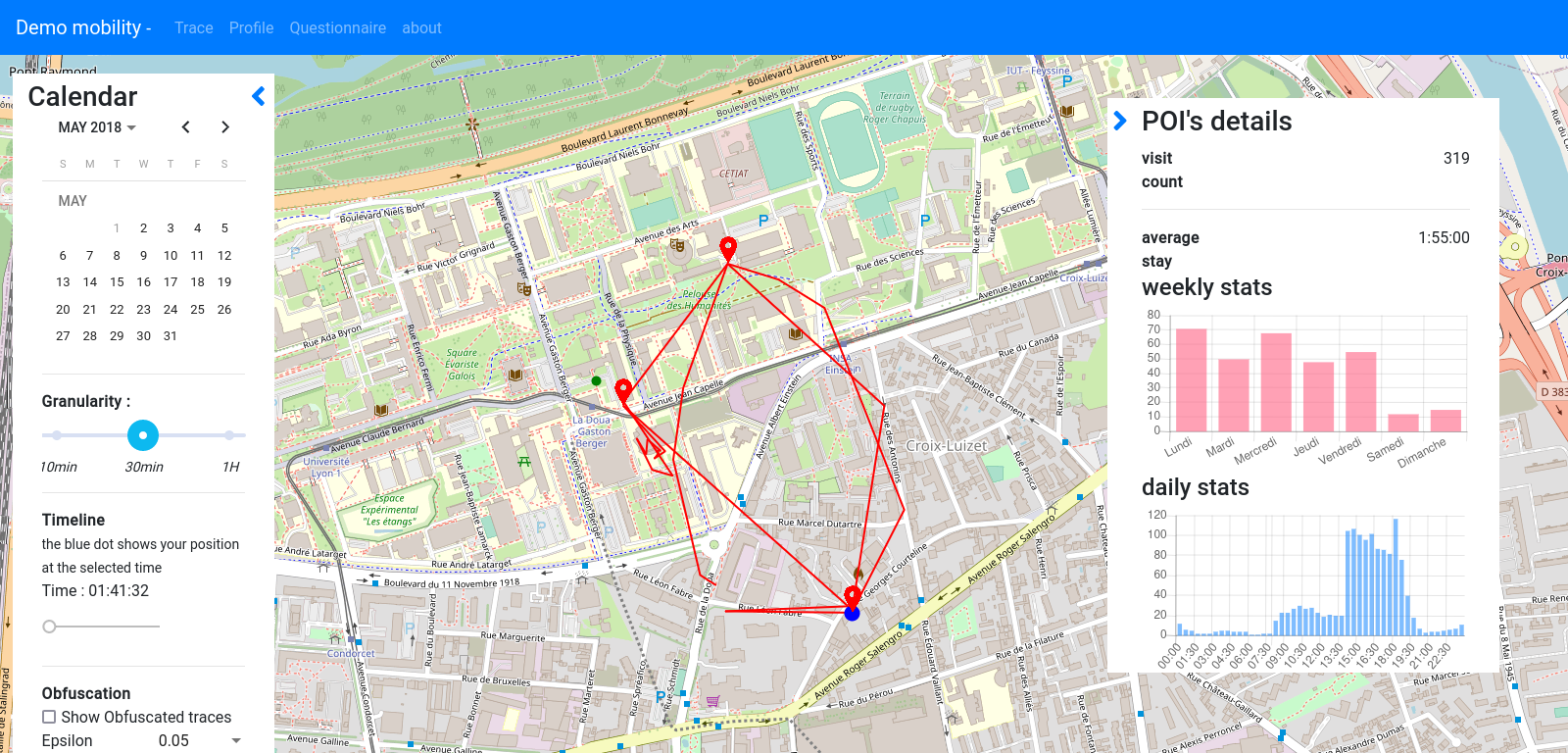}
    \caption{Screenshot of the demonstration's interface. Clicking on a POI yields attendance statistics.}
    \label{fig:traces}
\end{figure}

After the first behavior questionnaire, participants had to explore a demonstration platform informing them about privacy risks linked to location traces through example data. 
%This demonstration platform has been previously introduced in~\citet{boutet:hal-02421828}.
We use the~\cite{boutethal-02421828} demonstration platform.
% , and can be found at \url{https://preserve-1.inrialpes.fr/demo}.
%This demo either asks participants to export their own position history exported from Google Takeout and to import it on the platform, or provides the possibility of using example data.
%Participants were asked to export their own position history exported from Google Takeout and to import it on the platform. 
%They also had the possibility of using example data.
Participants were asked to use the platform to explore these location traces and to inspect the information that can be inferred.
%They were then able to explore the visualisation tool to inspect their location traces and proposed on the platform.
%\todo[inline]{Antoine: what do the traces precisely consist of? I would highlight that we're talking about smartphone traces.}
The demonstration typically offers to visualize one's traces per day (see Figure~\ref{fig:traces}), which can raise awareness about how easy it is to identify Points of Interests (POIs) -- such as the home, the workplace or any other visited places during the day --, and presents metadata (e.g., address, category of the place, attendance statistics).

\begin{figure}[t]
    \centering
    \vspace{2mm}
    \includegraphics[width=.44\textwidth]{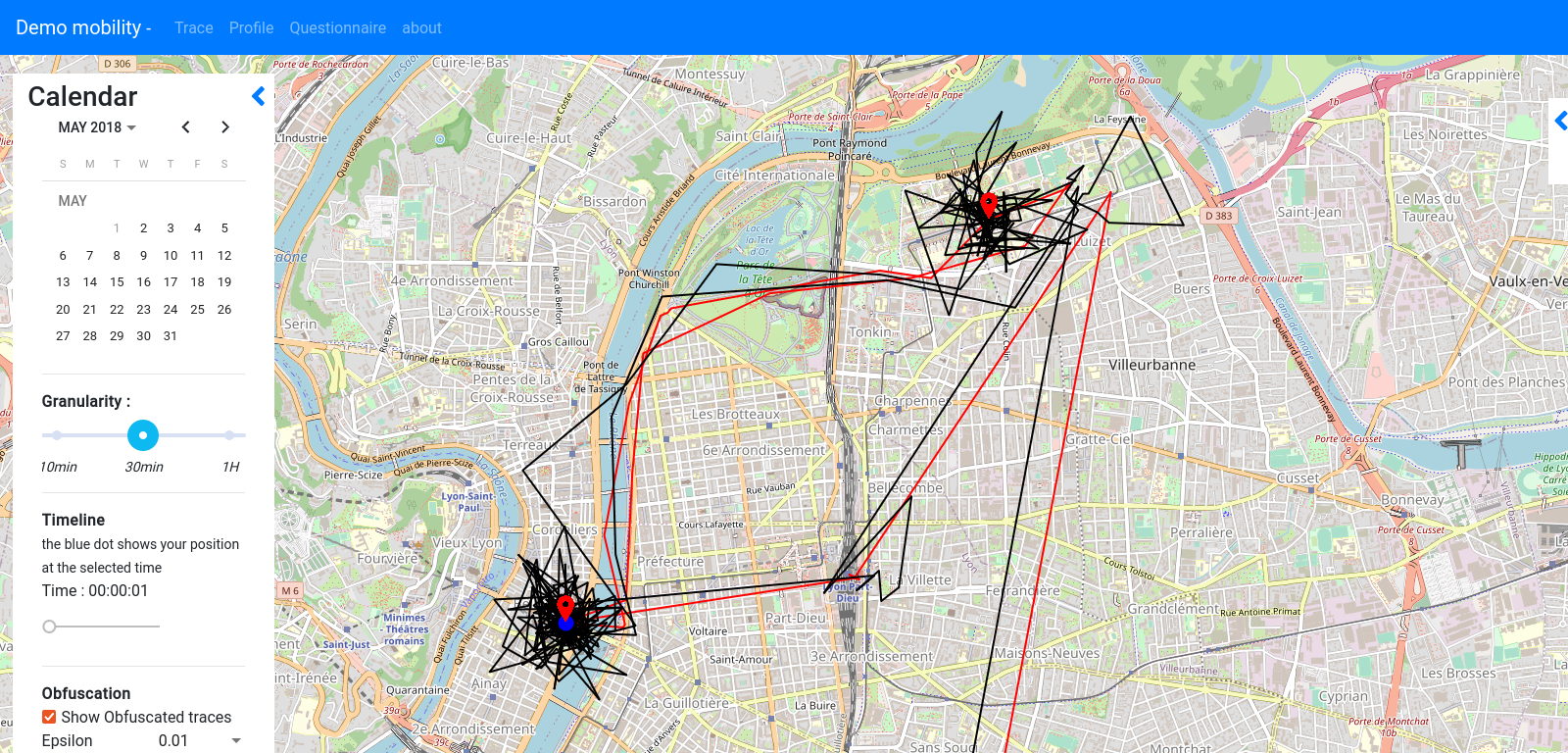}
    \caption{Illustration of the defense mechanism. Here the red lines show the initial traces, and the black lines represent the traces after the application of the noise.}
    \label{fig:obfuscated}
    \vspace{-2mm}
\end{figure}

The demonstration also offers a defense mechanism based on Differential Privacy~\cite{dwork_differential_2006} in the form of a slider, which controls the level of noise injected into the location data (see Figure~\ref{fig:obfuscated}).
This feature is presented as a means to stimulate interest in privacy protection, and to raise awareness of the degradation of data quality when we increase protection with greater noise (to showcase the utility and privacy trade-off). % between privacy and utility).

\begin{figure}[!ht]
    \centering
    \includegraphics[width=.44\textwidth]{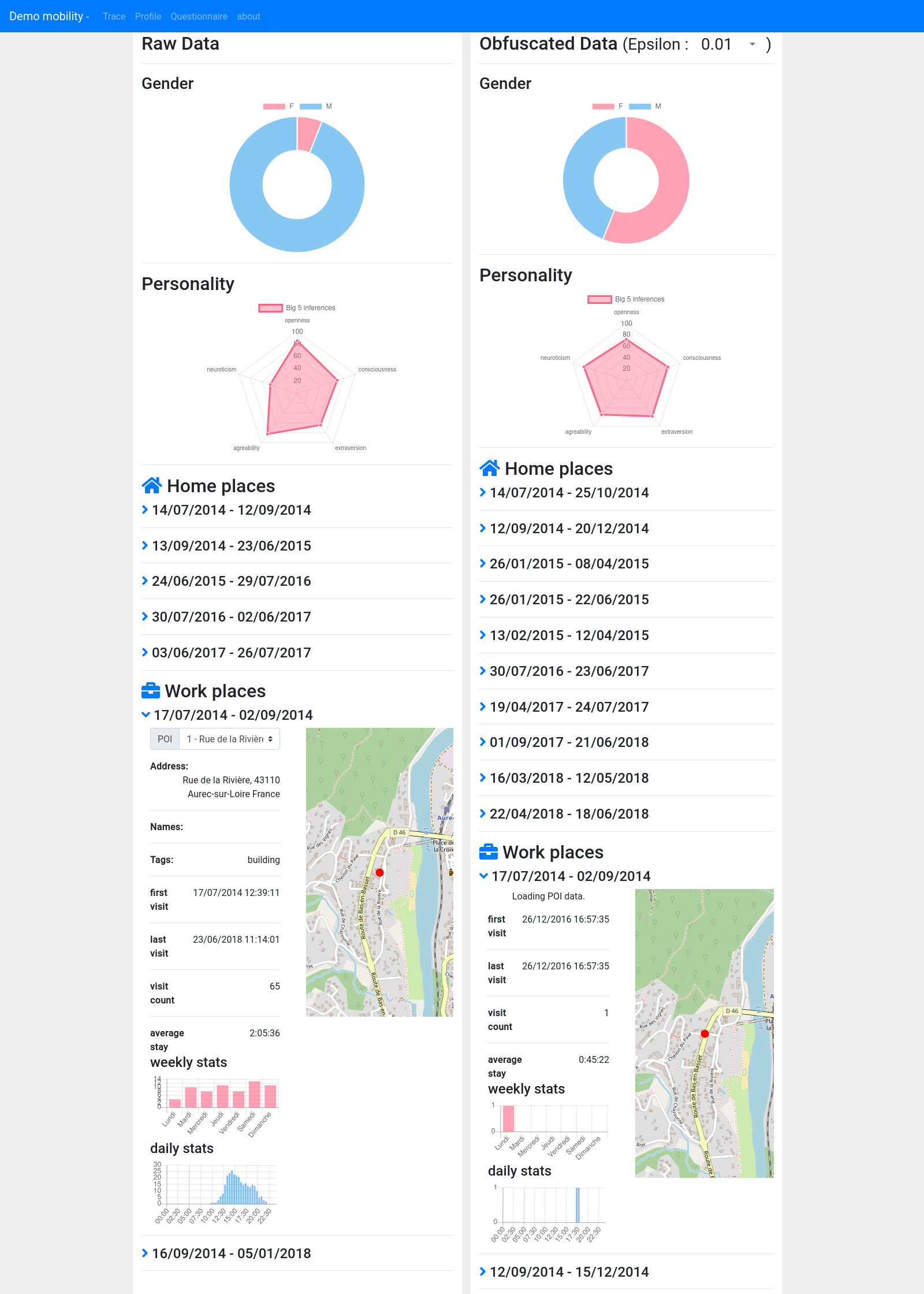}
    \caption{Inference performed by the platform. The left-hand side shows inference from raw data while the right-hand side presents the reduced risks after the application of the defense mechanism (i.e., geo-indistinguishability~\cite{Andr_s_2013}). The noisy version hardly yields the gender, smooths the personality traits, makes the daily and weekly attendance statistics unusable, and infers inaccurate POIs.}
    \label{fig:inference}
    \vspace{-3mm}
\end{figure}

Finally, the demonstration highlights the risks raised by analyzing location traces and inferring gender, big five personality traits, home, and workplace (see Figure~\ref{fig:inference}).
Given the large number of inferences of sensitive information from location data reported in the literature (e.g., people encountered, religious affiliation, sexual orientation, or health status~\cite{10.1145/1868470.1868479}), only the presentation of these inferences fits fairly well into a representative risk of targeting  (i.e., based on predictions computed on data collected and potentially exchanged between different parties). 
% \textbf{Note that this demonstration has been designed and developed by one of the author and has been presented in an international peer-reviewed conference}\footnote{Anonymized, we will provide the reference and the URL to share the tool once the paper is published}.

%\textbf{Did the authors deisgn and implement this tool themselves? If so, the paper should include some of those details. The authors might also want to describe if and how they plan to share the tool with the broader research community.}

% \todo[inline]{We might have to take down this citation for the submission (double-blind review process)}
% \cite{boutet:hal-02421828}

%\subsection{Procedure}
%\label{subsec:procedure}
%\todo[inline]{What do you mean by procedure?}

\subsection{Data analysis}
\label{sec:analysis}

We performed a thematic analysis on the free text answered in the study (see Tables~\ref{tab:harms} and~\ref{tab:gdpr} in Appendix~\ref{app:codebooks}) 
to analyse \textit{qualitative} data.
We took inspiration from the steps described in~\cite{braun_2006_method}:
% for the thematic analysis of the qualitative data obtained: 
\begin{enumerate}
    \item read the answers;
    \item code the sentences from the previous step;
    \item merge the codes in normalised codes for data analysis;
    \item regroup the normalised codes from the previous step in themes;
    \item review the themes and the related sentences to verify the homogeneity between them.
\end{enumerate}

The analysis was performed by two independent annotators, and a consensus was reached at the end of the process.
For most fields, a full thematic analysis was neither required nor necessarily desirable; curation and normalisation of the data was however necessary.
Indeed, stopping at the third step was enough for several fields, such as the reasons given to revoke location permission.
A full thematic analysis was performed on the answers to \textit{Q32} ``How do you think location data could be misused by a malicious person?'', the themes of which were loosely derived from~\cite{citron_privacy_2021} (Physical harms, Economic harms, Reputational harms, Informational integrity, Psychological harms), see Table~\ref{tab:harms} in Appendix~\ref{app:codebooks}.

We regrouped our codes for \textit{Q36} ``What are the measures imposed by the GDPR that reduce these risks?'' under general legal categories (Rights, Consent, Clear purpose and transparency, Misc.) for statistical purposes, see Table~\ref{tab:gdpr} in Appendix~\ref{app:codebooks}. The rationale behind Q36 was to evaluate participants’ knowledge about the GDPR.
% \newpage

\subsection{General Statistics}
\label{sec:general}

Our participants are mostly equipped with Android phones: 60\% against 40\% for iOS users (\textit{Q2}). 
Our population is not gender balanced, our sample was composed of 25\% of women, 72\% of men, and 3\% prefer not to answer (\textit{Q26}). 
Participants are young with an average age around 21 with a small standard deviation ($\sigma=1.04$) (\textit{Q27}).
%\todo[inline]{My data is buggy, I can't read the age.}

\begin{figure}[t]
    \centering
    \vspace{-4mm}
    \includegraphics[width=.33\textwidth]{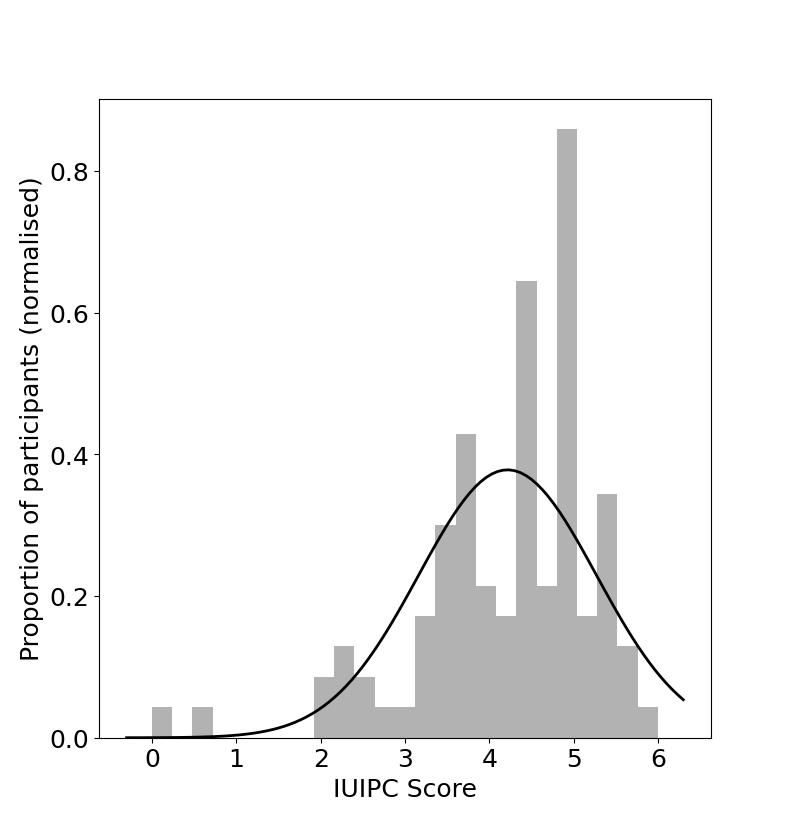}
    \vspace{-1mm}
    \caption{Plot of the Internet User’s Information Privacy Concerns (IUIPC) distribution of participants.}
    \label{fig:IUIPC}
    \vspace{-2mm}
\end{figure}

Their privacy concerns have been assessed using the Internet Users' Information Privacy Concerns (IUIPC) Score. 
The IUIPC is a scale widely used in privacy research, it reflects Internet users’ concerns about information privacy with a focus on ``individuals’ perceptions of fairness/justice in the context of information privacy''.
The distribution of IUIPC is depicted in Figure~\ref{fig:IUIPC}. 

Note that our population is within average with respect to their privacy concerns ($\mu$ = 4.2, $\sigma$ = 1.05)\footnote{See \cite{DBLP:journals/popets/ZuffereyNHH23}, for an example of less concerned participants ($\mu$ = 3.5, $\sigma$ = 1.6) and~\cite{naeini_privacy_2017} for more concerned participants with $\mu$ between 4.79 and 5.44 normalized to a [0-6] scale.} in spite of young adults being generally less concerned about their privacy than older populations (as mentioned in Section~\ref{sec:related}).

\section{Results}
\label{sec:results}
%\todo[inline]{Our weakness is that some results might be only remotely connected to the research questions, at least in the current state of the paper.}
% \todo[inline]{Refer all results to questions (now numbered in the app.).}

This section showcases the most relevant results for our research questions, with the first two subsections dedicated to answering RQ1 (``What are the perceptions and the understanding of young users’ privacy and its protection in a mobility context?''), and the last subsection more %specifically 
tailored to RQ2 (``What is the impact of visualization of location traces and associated privacy risks on these concepts?'').

\subsection{Privacy risky practices}
\label{sec:RPractices}

Participants do not adopt safe practices in terms of privacy. Specifically, they tend to underestimate the number of apps that have access to the location, and they tend not to think about disabling location access for apps that are not actively or forget to do so. 

\begin{figure}[!h]
    \centering
    \vspace{-1mm}
    \includegraphics[width=0.44\textwidth]{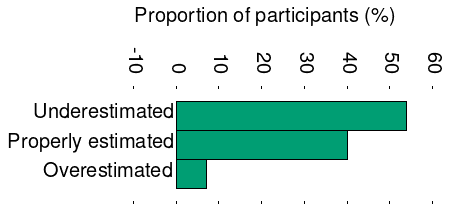}
    \vspace{-5mm}
    \caption{Participants tend to underestimate the number of apps that have access to the location (\textit{Q3} against \textit{Q8}).}
    \label{fig:estimation}
\end{figure}

\noindent\textbf{Underestimation of the number of apps that have access to location.}
%\subsection{Data Sharing behavior}
%\label{sec:Rdatasharing}
%We first note that 13\% of the participants do not know where to see which apps are tracking their location (\textit{Q5}), which means that this 1/8th of our participants are not aware of how to control their own privacy settings.
%Another significant finding is that 67\% allow location tracking (whether location services are enabled) (\textit{Q6}).
%% \textbf{Q6: "Please open the setting of your phone and check if you
%% allow location tracking. (Enabled/Disabled?)" %Does this mean that location
%% services are enabled? Or does it also mean that %location history is saved
%% (e.g., [Google Location History](https://support.google.com/android/answer/3118687?hl=en)
%% or [iOS Significant Locations](https://support.apple.com/guide/iphone/delete-significant-locations-iph32b15b22f/ios)
%% are enabled)?
%% }
%Amongst these 65 participants, almost all of them (61) had activated location enhancement via Wi-Fi, and 38 through Bluetooth (sometimes in addition to Wi-Fi) (\textit{Q7}).
%\todo[inline]{What is 'location tracking' BTW? I need to know precisely to elaborate the discussion on this point.}
%
%
%
%
%
%\subsection{Underestimation of the number of apps that have access to location}
Participants were asked to estimate how many mobile apps have access to the location on their smartphone ``on top of their head'' (\textit{Q3}), and this data was then compared to the actual number (both for ``always'' and for ``when using the app'') (\textit{Q8}).
Unsurprisingly, the majority of participants (52) underestimated the number of apps with location access, 7 overestimated it, but 38 were correct (with 17 on the open range, i.e., they selected the maximum value), as presented in Figure~\ref{fig:estimation}.
We interpret this result as a wrongful mental model due to 1) the lack of transparency of apps regarding their data collection, 2) an ever-increasing number of applications, combined with 3) a lack of a central controlling system for privacy permissions.

While the average number of apps with continuous access to the location was pretty low but with a high standard deviation ($\mu$ = 2.74, $\sigma$ = 6.32) -- meaning that most participants do not allow continuous access to their location except for a few outliers with a high number of apps --, we note that a much higher number of apps had access to location when on use ($\mu$ = 11.04, $\sigma$ = 8.93)(\textit{Q8}).

Amongst the applications with access to location (\textit{Q9}), Google Maps is leading the list ``on top of their heads'' (38 mentions), but has more access in practice (15 for ``always'' and 27 ``when using the app''); it is followed by Instagram (respectively 27, 4, 31) and Snapchat (26, 4, 22).

%\paragraph{Underestimation of the number of apps}

\begin{figure}[!h]
    \centering
    \includegraphics[width=0.42\textwidth]{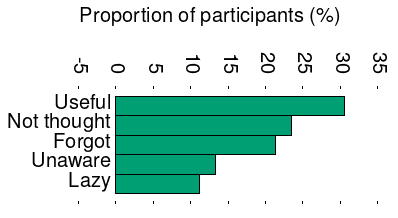}
    \vspace{-1mm}
    \caption{Participants tend not to think about disabling location access for apps that are not actively used or forget to do so (\textit{Q11}).}
    \vspace{-2mm}
    \label{fig:forget}
\end{figure}

\noindent\textbf{Participants tend to forget about data-sharing.}
%
%\subsection{Participants tend to forget about data-sharing}
%\label{sec:forget}
% When asked why they were leaving access to location permission to apps not actively used, 12 answered hey hey  regarde ce que je viens de marquer si ça répond à ce point ;)
%
The results show that most participants had installed numerous applications with access to location (\textit{Q8}).
In addition, we see that 74.2\% (46 actively use most, 26 only some of them) have at least one application not actively used but with current access to location data (\textit{Q10})\footnote{We orally clarified to participants that ``actively used'' amounts to a weekly use or more frequent.}.
When asked why these unused applications still have access to their location (\textit{Q11}), a majority of the participants have either forgotten (11) or not thought (12) about removing access (see Figure~\ref{fig:forget}).
Another proportion of participants are either lazy (6) or unaware (7) that these applications collect location data. 
Finally, the rest of the participants (15) prefer to leave the permission to the app in case they need to use it.

Basically, participants have difficulties remembering their apps' privacy practices by themselves, which make the latest features of iOS (``Offload Unused Apps'') and Android (``Unused Apps'') all the more relevant.

\noindent\textbf{Authorisation and Control.}
We asked participants questions regarding the control of location access to mobile applications (\textit{Q23}). % presented in Figure~\ref{fig:habit}).
Only 4\% think it is not possible to revoke (cancel) previously granted mobile app access (\textit{Q18}) and 16\% of the participants never modified the privacy settings related to their location data (\textit{Q19}).
A greater proportion of participants reduced the availability of their location data than those who increased their availability (75\% versus 8\%).
For those who said they had already revoked access, the reasons given by 68\% of participants felt that this access was not necessary for the application (``I'm not for sale''); or did not understand why this access was necessary (\textit{Q21}).

\subsection{Participants are poorly aware of the risks}

They do not know how location can be captured, they are not aware of the great inference capacity associated with the analysis of location data, and they are unable to cite cases of personal data leaks or scandals related to their use despite increased media coverage.

\begin{figure}[!h]
    \centering
    %\vspace{-2mm}
    \includegraphics[width=0.41\textwidth]{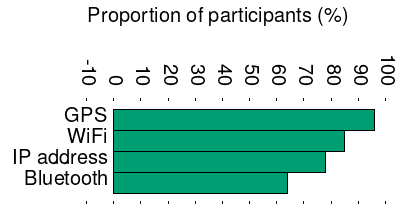}
    \vspace{-1mm}
    \caption{Proportion of participants aware of source of data from which location can be inferred (\textit{Q12}).}
    \label{fig:techo}
    %\vspace{-2mm}
\end{figure}

\noindent\textbf{Participants have misconceptions about how location can be captured.}
%
%Plot: GPS 96\%, WiFi 85\%, Bluetooth 64\%, IP address 78\%
% \todo[inline]{Plot needed}
To assess the perception of privacy risks linked to the sharing of location information, we asked the participants how a mobile application could capture location (\textit{Q12}).
The responses are displayed in Figure~\ref{fig:techo} and show that not all participants are aware of the localization capabilities of the main technologies embedded in a phone.
Although GPS is identified by almost all participants (96\%), we see that only 85\%, 64\% and 78\% of participants are aware that respectively WiFi, Bluetooth, and IP addresses can be used as proxies to infer location.
% Strange because these are the means reported in the main location APIs -> no mobile development or other?

\begin{figure}[!h]
    \centering
    \includegraphics[width=0.48\textwidth]{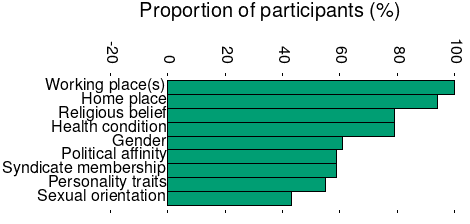}
    \vspace{-4mm}
    \caption{Answers to \textit{Q13} ``Do you think it is possible to infer the following information from the location?''.}
    \label{fig:risk}
\end{figure}

\noindent\textbf{Unaware of the inference capacity from location data.}
%
%
%Working place(s)		100\%
%Home place		94\%
%Religious belief		79\%
%Political affinity		59\%
%Personality traits		55\%
%Gender		61\%
%Sexual orientation		43\%
%Health condition		79\%
%Syndicate membership		59\%
% \todo[inline]{Plot needed}
In the behavior questionnaire before the demonstration, we then asked participants to identify what information could be inferred from location amongst a list of information generally categorised as sensitive (\textit{Q13})\footnote{Most of these categories are listed as special categories of data by Article 9(1) of the GDPR, see Section~\ref{sec:relatedPrivacy}.}.
The results are represented in Figure~\ref{fig:risk} and show that the risks of inference are moderately known.
%\todo[inline]{Support last claim with ref to what is actually possible to infer.}
Although all participants perceived that the place of work can be inferred from location traces, only 94\% think that the place of residence can also be inferred. 
These inferences are based on the time spent in a location over office hours and overnight stays. 
This change in perception raises questions about the rationality or basis for the response because we spend more time in the home than in the workplace.
% Then we see that this percentage drops when it comes to more precise inference linked to the observation of visits to certain places.
Finally, we observe that 10\% of participants do not believe that this information can lead to re-identification (\textit{Q15}), which is a surprisingly high number considering that all participants had previously followed courses on data science and AI.
On the other hand, when asked why a service provider might be interested in the location of individuals (\textit{Q14}), 88\% responded that this information is for marketing purposes. 
So they still have the perception that this information could be used for targeting or profiling.

\noindent\textbf{Little echo about data leaks or privacy scandals.}
%
% \todo[inline]{Tune down subsection's title}
The majority of participants are unable to list cases of personal data leaks or scandals linked to their uses despite press articles regularly published on the subject (\textit{Q33}).
In view of the numerous scandals and cases linked to personal data reported very regularly by the press (e.g., ransomware and data leaks in hospitals, attempted influence on social networks), this lack of awareness of the risks raises questions about their exposure to the media.
Only 32 knew about a scandal, half of these cases were Cambridge Analytica (16 mentions), while major scandals such as Pegasus or the Snowden leaks were barely mentioned (3 mentions each).
The medium-high reported IUIPC of the participants (see Section~\ref{sec:general}) does not seem to come from exposure to cases concerning privacy violations reported by the press.
%\todo[inline]{How do we make sense of this data against a high reported IUIPC? Could seem contradictory.}

\subsection{The risk demonstrator increased awareness of participants about privacy}
%\subsection{On the impact of visualization of location traces}
\label{sec:impact}

After having experimented with the demonstration, a third of the participants (33) were planning on limiting the access to mobile privacy permissions, and deleting the history of mobility on the Google Takeout platform (\textit{Q37}).
%As mentioned before, 51.5\% declared to plan using PETs after having experimented with the demonstration platform (\textit{Q38}).
We contend that the visualisation of location traces and the risks associated with the inference increased awareness of participants towards privacy, and encouraged them to adopt more privacy-preserving behaviors.

\noindent\textbf{Relative surprise for the number of possible inferences.}
Once the participants have been able to analyze location traces and discover the information that can be learned with the demonstrator, more than half of the participants (56\%) declared being surprised by the number of inferences made possible through location data (\textit{Q29}). 
Additionally, when asked whether they feel well protected against the various threats observed, a large majority of participants responded that they felt unprotected (84\%) (\textit{Q35}).
Knowing this inference ability, participants were asked if they think Google analyzes users' location data (\textit{Q30}), and if they think Google shares this data with third parties (\textit{Q31}).
The results show that 93\% of participants believed that Google analyzes their location data and 89\% believed that their data is shared with third parties.
Even after the inspection of location traces with the platform, 7\% of participants still did not think that location data can be used to re-identify users (\textit{Q34}).
%re identification est à 7\%

%THEMES (loosely based on Solove’s taxonomy)	Codes belonging to the theme	Comments
%Physical harms	Physical harm, harassment	Mentioned (11+20=31 mentions)
%Economic harms	Blackmail, robbery, scam, identity theft	Prevalent (9+28+6+8=51 mentions)
%Reputational harms	Doxing	Anecdotical (3 mentions)
%Informational integrity	Surveillance, stalking, inference, data selling, targeted advertising	Very prevalent (40+17+22+6+3=88 mentions)
%Psychological harms	Manipulation	Occasionally mentioned (8 mentions)
%
%\todo[inline]{Plot needed}
%

\noindent\textbf{Worries about surveillance, economic, and physical harms.}
The most worrisome inferences are informational harms (surveillance, stalking\footnote{Here, \textit{stalking} refers to close surveillance without intervention, we coded surveillance with physical and/or oral intervention under \textit{harassment}.}, inference, data selling, targeted advertising; 88 mentions), followed by economic harms (blackmail, robbery, scam, identity theft; 51 mentions), and finally physical harms (physical harm, harassment; 31 mentions) (\textit{Q32}).

\begin{figure}[!h]
    \centering
    \includegraphics[width=0.48\textwidth]{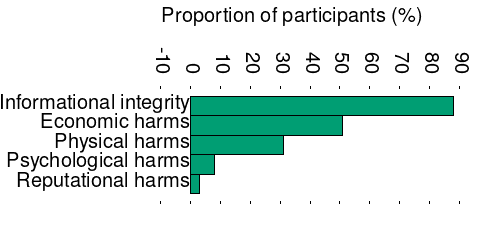}
    \vspace{-8mm}
    \caption{The most worrisome inferences are informational harms (i.e., surveillance, stalking, inference, data selling, targeted advertising) (\textit{Q32}).}
    \label{fig:harms}
\end{figure}

% See Table~\ref{tab:harms} for the codebook.
%
We observed interesting preconceived ideas regarding those harms.
% It is interesting to highlight a few interpretations. 
First, in spite of being in the most cited category, targeted advertising was specifically mentioned only three times, whereas it is the main privacy-invasive goal of location data analysis by Google (from which the traces were extracted). 
Second, robbery was over-represented in the answers (28 mentions, 2nd most cited code) although it does not happen so often in practice\footnote{A possible explanation can be that of the black swan theory, or the presence of websites raising awareness of this specific risk such as~\url{https://pleaserobme.com/}}.

\noindent\textbf{Positive attitude towards protective mechanisms.}
%
% \todo[inline]{Ensure non-repetition with items in section~\ref{sec:impact}}
%\paragraph{PETs}
Despite the lack of awareness about the risks, participants are inclined to use protection tools. 
When asked whether technologies could reduce and provide greater control over personal information disclosed via location data, 76.3\% of participants responded that they would be willing to use Privacy Enhancing Technologies (PETs) (\textit{Q16}). 
Consistently, 51.5\% declared to plan using PETs after having experimented with the demonstration platform (\textit{Q38}). % (see Figure\ref{fig:use_pets}).
However, a lower score on question Q38 than on question Q16 means that participants do not have a concrete idea of existing privacy-enhancing tools.
On the other hand, 84\% of participants said they have never looked to see if this type of technology or tools existed (\textit{Q39}).

\section{Discussion}
\label{sec:discussion}
This section formulates practical recommendations and highlights the limitations of our work.

\subsection{Recommendations}
We craft in this section recommendations for the design of better privacy-friendly systems, and of Transparency Enhancing Technologies (TETs) to foster privacy awareness.
These recommendations are drawn from both participants' insights and our own interpretations.
% of the results.

% \todo[inline]{Mettre en exergue la difficulté qu'ont les utilisateurs à gérer manuellement leurs paramètres de localisation}
\textbf{Assist in the revocation of apps.}
Our results show that participants tend to forget about data-sharing, %(see Section~\ref{sec:RPractices}), 
where more than a quarter of the participants declared not thinking about revoking access (12), forgot to do it (11), or are simply too lazy (6), and that they have inaccurate representations of the number of apps with access to their location (see Section~\ref{sec:RPractices}). %sec:Rdatasharing}).
However, we note that 14 participants suggested including a system-wide reminder of which apps can access location as recommendations for better and more usable Privacy Enhancing Technologies (PETs). 
Note that the latest version of Android and iOS indicate when an app accesses location, and Android now offers a centralized interface indicating which apps require and use geolocation.
This centralized interface can be hard to find amongst the different settings, and no participant were aware of its existence.
However, both Android and iOS lack a centralized interface to inform which app is \textit{currently} accessing location, nor do they provide fine-grain data regarding its use (i.e., how many times did each app request geolocation).
We therefore recommend that \textbf{mobile OSs integrate a centralised and easily accessible interface about location usage}, providing not only which geolocation permission has been granted to each app but also a fine-grain data regarding its uses.

\textbf{More transparency.}
Our results also show that knowing which apps have access to location is not enough; 16 participants also proposed more transparency and more specifically to be able to be informed at any time of which applications are collecting data.
Seeing the lack of perception of privacy risks and the impact of tools allowing to calculate and visualize these risks on this perception, we suggest improving transparency by \textbf{providing users with statistics and metrics of risk} associated with data collection for each application. 
We believe that this personalized risk analysis would be an important lever to help users to evaluate the trade-off between utility and respect for their privacy and to better calibrate what they wish to share.

\textbf{Access privacy permissions from the app.}
Another highlight from the participants' answers is the possibility to access privacy permissions directly from within the app.
Indeed, today's privacy permissions (on both iOS and Android) are mainly available through the general settings of the smartphone, which can be hard to access for lay users\footnote{Note that it is possible to access privacy permissions for an app from some Android launchers, but \textbf{not} from within the app.}.
We therefore recommend mobile OSs to \textbf{include a shortcut to an app's privacy permissions from the app itself}, for instance from the status bar or by executing a special gesture.
This suggested feature is currently unavailable on modern mobile operating systems and has not been suggested anywhere else to the best of our knowledge.

\subsection{Limitations}
\label{sec:limitations}
First, our demographics do not represent a fair sample of digital users.
Our participants were mostly males, all of them young adults educated in a top European engineering school, although with different backgrounds and technical skills.
This bias in our representativeness played a part in the socio-economic backgrounds of participants, who are most likely to come from well-off social environments.
Recruiting participants in universities often faces this limitation in academic studies on usages. %~\cite{ddcc67d7ea224a9ebe46ab14be1f8600, 10.1145/3372296, 10.1145/2702123.2702210, 274429}.
For instance, \cite{ddcc67d7ea224a9ebe46ab14be1f8600} involved 40 participants mostly young and well-educated, while \cite{10.1145/3372296} involved 80 participants, mostly young, % of them were %quite young, 
with 45\% in the 18–25 age group and 20\% between the ages of 26 and 35, a majority ($n=54$) were bachelor students or had their bachelor degrees. \cite{almuhimedi2015your} involved 23 participants (65\% female; ages 18–44, median=23), and \cite{274429} involved 392 participants ranging from 25 to 35.

The transferability of the result to the rest of the population therefore remains an open question.
However, given that our sample is predominantly composed of ``digital natives'' -- who are often assumed to be more comfortable with personal information sharing and more knowledgeable about digital technologies --, we expected a better perception about the privacy risk than amongst other age groups. 
This expectation also has to be put in the perspective of their curriculum: students in computer engineering with academic knowledge of data science, AI, and rudiments of GDPR.
%We will clarify this point and better motivate why our study should be published even with a small, non-representative sample of users.

Second, the scope of our work must be restricted to a specific type of demonstrator, % tool, 
our results should not be generalized to all visualization and awareness tools. 
Moreover, the results largely presented self-reported and planned behaviors.
Therefore, an interesting research avenue for future work is to study the \textit{actual behavioral} changes introduced by the experiment of the tool, and long-term behavior analysis through \textit{longitudinal studies}.
% \textcolor{red}{and long-term behavior analysis}.
Perceptions can also change according to the type of apps and their access to location data, as well as other sensitive personal data collected.
% by mobile apps.
% \textcolor{red}{as well as other sensitive personal data collected by mobile apps}.

Third, participants used example data on the demonstration platform, which is perhaps less illustrative than if they had inspected their own location data. We envision working with personalized examples in the future.
%Similarly, 
Our results also lack perspective on whether our participants were able to deeply understand the goal and the meaning of obfuscated traces and the epsilon number (see~\cite{karegar2022exploring}).

%\todo[inline]{Third, they only used example data on the demo platform, so less illustrative}

\section{Conclusion} % and future work}
\label{sec:conclusion}

Through a survey involving $n=99$ young mobile phone users, this work contributes to a better understanding of the practices in relation to the sharing of their location data with mobile applications as well as the associated privacy issues.
By qualitatively and quantitatively analyzing user perceptions and self-reported sharing behaviors, we provide valuable insights for researchers and privacy practitioners to better understand users and better design new Privacy Enhancing Technologies (PETs) and Transparency Enhancing Technologies (TETs) related to location data.
We also showed the importance of displaying a more %personalized and 
interactive risk analysis to make users better aware of privacy risks.

\section*{Acknowledgements}

This work has been supported by the ANR 22-PECY-0002 IPOP (Interdisciplinary Project on Privacy) project of the Cybersecurity PEPR, and partially supported by the Wallenberg AI, Autonomous Systems and Software Program (WASP) funded by the Knut and Alice Wallenberg Foundation.
We would like to thank Farzaneh Karegar and Nataliia Bielova for their feedback on earlier drafts, and anonymous reviewers who helped us improve this article.

\bibliographystyle{plain}
\bibliography{references}

\begin{thebibliography}{51}
\providecommand{\natexlab}[1]{#1}

\bibitem[{Achara et~al.(2013)Achara, Castelluccia, Lefruit, Roca, Baudot, and Delcroix}]{achara:hal-00917417}
Achara, J.~P.; Castelluccia, C.; Lefruit, J.-D.; Roca, V.; Baudot, F.; and Delcroix, G. 2013.
\newblock {Mobilitics: Analyzing Privacy Leaks in Smartphones}.
\newblock \emph{{ERCIM News}}.

\bibitem[{Almuhimedi et~al.(2015)Almuhimedi, Schaub, Sadeh, Adjerid, Acquisti, Gluck, Cranor, and Agarwal}]{almuhimedi2015your}
Almuhimedi, H.; Schaub, F.; Sadeh, N.; Adjerid, I.; Acquisti, A.; Gluck, J.; Cranor, L.~F.; and Agarwal, Y. 2015.
\newblock Your location has been shared 5,398 times! A field study on mobile app privacy nudging.
\newblock In \emph{CHI}, 787--796.

\bibitem[{Andreas et~al.(2016)Andreas, Hugo, Tobias, Konrad, and Felix}]{journals/popets/Kurtz16}
Andreas, K.; Hugo, G.; Tobias, B.; Konrad, R.; and Felix, F. 2016.
\newblock Fingerprinting Mobile Devices Using Personalized Configurations.
\newblock \emph{PoPETs}, 2016(1): 4--19.

\bibitem[{Andr{\'{e} }s et~al.(2013)Andr{\'{e} }s, Bordenabe, Chatzikokolakis, and Palamidessi}]{Andr_s_2013}
Andr{\'{e} }s, M.~E.; Bordenabe, N.~E.; Chatzikokolakis, K.; and Palamidessi, C. 2013.
\newblock Geo-indistinguishability.
\newblock In \emph{CCS}.

\bibitem[{Balash et~al.(2022)Balash, Wu, Grant, Reyes, and Aviv}]{277130}
Balash, D.~G.; Wu, X.; Grant, M.; Reyes, I.; and Aviv, A.~J. 2022.
\newblock Security and Privacy Perceptions of {Third-Party} Application Access for Google Accounts.
\newblock In \emph{USENIX Security}, 3397--3414.

\bibitem[{Barbaro and Zeller(2006)}]{article2}
Barbaro, M.; and Zeller, T. 2006.
\newblock A Face is exposed for AOL searcher no. 4417749.
\newblock \emph{New York Times}.

\bibitem[{Barth et~al.(2019)Barth, {de Jong}, Junger, Hartel, and Roppelt}]{BARTH201955}
Barth, S.; {de Jong}, M.~D.; Junger, M.; Hartel, P.~H.; and Roppelt, J.~C. 2019.
\newblock Putting the privacy paradox to the test: Online privacy and security behaviors among users with technical knowledge, privacy awareness, and financial resources.
\newblock \emph{Telematics and Informatics}, 41: 55--69.

\bibitem[{Barth-Jones(2012)}]{article}
Barth-Jones, D. 2012.
\newblock The 'Re-Identification' of Governor William Weld's Medical Information: A Critical Re-Examination of Health Data Identification Risks and Privacy Protections, Then and Now.
\newblock \emph{SSRN Electronic Journal}.

\bibitem[{Beckwith(2003)}]{1203752}
Beckwith, R. 2003.
\newblock Designing for ubiquity: the perception of privacy.
\newblock \emph{IEEE Pervasive Computing}, 2(2): 40--46.

\bibitem[{Bielova et~al.(2024)Bielova, Chammat, Toubiana, Hary, Nguyen, and Litvine}]{bielova:hal-04235032}
Bielova, N.; Chammat, M.; Toubiana, V.; Hary, E.; Nguyen, A.; and Litvine, L. 2024.
\newblock {The Effect of Design Patterns on (Present and Future) Cookie Consent Decisions: Supplemental Materials}.
\newblock In \emph{USENIX Security}.

\bibitem[{Boutet and Gambs(2019)}]{boutet:hal-02421828}
Boutet, A.; and Gambs, S. 2019.
\newblock {Demo: Inspect what your location history reveals about you Raising user awareness on privacy threats associated with disclosing his location data}.
\newblock In \emph{{CIKM}}, 2861--2864.

\bibitem[{Braun and Clarke(2006)}]{braun_2006_method}
Braun, V.; and Clarke, V. 2006.
\newblock Using thematic analysis in psychology.
\newblock \emph{Qualitative Research in Psychology}, 3(2): 77--101.

\bibitem[{Citron and Solove(2021)}]{citron_privacy_2021}
Citron, D.~K.; and Solove, D.~J. 2021.
\newblock Privacy {Harms}.
\newblock \emph{SSRN Electronic Journal}.

\bibitem[{De~Montjoye et~al.(2013)De~Montjoye, Hidalgo, Verleysen, and Blondel}]{de2013unique}
De~Montjoye, Y.-A.; Hidalgo, C.~A.; Verleysen, M.; and Blondel, V.~D. 2013.
\newblock Unique in the crowd: The privacy bounds of human mobility.
\newblock \emph{Scientific reports}, 3(1): 1--5.

\bibitem[{Debatin et~al.(2009)Debatin, Lovejoy, Horn, and Hughes}]{10.1111/j.1083-6101.2009.01494.x}
Debatin, B.; Lovejoy, J.~P.; Horn, A.-K.; and Hughes, B.~N. 2009.
\newblock {Facebook and Online Privacy: Attitudes, Behaviors, and Unintended Consequences}.
\newblock \emph{Journal of Computer-Mediated Communication}, 15(1): 83--108.

\bibitem[{Dwork(2006)}]{dwork_differential_2006}
Dwork, C. 2006.
\newblock Differential {Privacy}.
\newblock In \emph{Automata, {Languages} and {Programming}}, volume 4052, 1--12. Berlin, Heidelberg: Springer Berlin Heidelberg.

\bibitem[{Eckersley(2010)}]{eckersley2010unique}
Eckersley, P. 2010.
\newblock How unique is your web browser?
\newblock In \emph{PETS}, 1--18.

\bibitem[{Englehardt and Narayanan(2016)}]{10.1145/2976749.2978313}
Englehardt, S.; and Narayanan, A. 2016.
\newblock Online Tracking: A 1-Million-Site Measurement and Analysis.
\newblock In \emph{CCS}, 1388–1401.

\bibitem[{Farke et~al.(2021)Farke, Balash, Golla, D{\"u}rmuth, and Aviv}]{274582}
Farke, F.~M.; Balash, D.~G.; Golla, M.; D{\"u}rmuth, M.; and Aviv, A.~J. 2021.
\newblock Are Privacy Dashboards Good for End Users? Evaluating User Perceptions and Reactions to Google{\textquoteright}s My Activity.
\newblock In \emph{USENIX Security}, 483--500.

\bibitem[{{FORCE11}(2020)}]{fair}
{FORCE11}. 2020.
\newblock The FAIR Data principles.
\newblock \url{https://force11.org/info/the-fair-data-principles/}.

\bibitem[{Gamarra et~al.(2019)Gamarra, Meri{\~n}o~Fuentes, Calabria~Sarmiento, Gutierrez~Acosta, Barrios~Barrios, Leal, and Wightman~Rojas}]{gamarrapercepcion}
Gamarra, M.; Meri{\~n}o~Fuentes, I.; Calabria~Sarmiento, J.; Gutierrez~Acosta, O.; Barrios~Barrios, M.; Leal, N.; and Wightman~Rojas, P.~M. 2019.
\newblock Privacy Perception in Location-Based Services for Mobile Devices in the University Community of the North Coast of Colombia.
\newblock 23(1).

\bibitem[{Gambs, Killijian, and del Prado~Cortez(2010)}]{10.1145/1868470.1868479}
Gambs, S.; Killijian, M.-O.; and del Prado~Cortez, M. N.~n. 2010.
\newblock Show Me How You Move and I Will Tell You Who You Are.
\newblock In \emph{SIGSPATIAL International Workshop on Security and Privacy in GIS and LBS}, 34–41.

\bibitem[{Gebru et~al.(2021)Gebru, Morgenstern, Vecchione, Vaughan, Wallach, Iii, and Crawford}]{gebru2021datasheets}
Gebru, T.; Morgenstern, J.; Vecchione, B.; Vaughan, J.~W.; Wallach, H.; Iii, H.~D.; and Crawford, K. 2021.
\newblock Datasheets for datasets.
\newblock \emph{Communications of the ACM}, 64(12): 86--92.

\bibitem[{Islami, Fischer-Hübner, and Papadimitratos(2022)}]{islami_capturing_2022}
Islami, L.; Fischer-Hübner, S.; and Papadimitratos, P. 2022.
\newblock Capturing drivers’ privacy preferences for intelligent transportation systems: {An} intercultural perspective.
\newblock \emph{Computers \& Security}, 123: 102913.

\bibitem[{Kang and Jung(2021)}]{kang2021smart}
Kang, H.; and Jung, E.~H. 2021.
\newblock The smart wearables-privacy paradox: A cluster analysis of smartwatch users.
\newblock \emph{Behaviour \& Information Technology}, 40(16): 1755--1768.

\bibitem[{Karegar, Alaqra, and Fischer-H{\"u}bner(2022)}]{karegar2022exploring}
Karegar, F.; Alaqra, A.~S.; and Fischer-H{\"u}bner, S. 2022.
\newblock Exploring $\{$User-Suitable$\}$ metaphors for differentially private data analyses.
\newblock In \emph{SOUPS}, 175--193.

\bibitem[{Karegar, Pettersson, and Fischer-H\"{u}bner(2020)}]{10.1145/3372296}
Karegar, F.; Pettersson, J.~S.; and Fischer-H\"{u}bner, S. 2020.
\newblock The Dilemma of User Engagement in Privacy Notices: Effects of Interaction Modes and Habituation on User Attention.
\newblock \emph{ACM Trans. Priv. Secur.}, 23(1).

\bibitem[{Kaushik et~al.(2021)Kaushik, Yao, Dewitte, and Wang}]{274429}
Kaushik, S.; Yao, Y.; Dewitte, P.; and Wang, Y. 2021.
\newblock "How I Know For Sure": People{\textquoteright}s Perspectives on Solely Automated {Decision-Making} ({{{{{SADM}}}}}).
\newblock In \emph{SOUPS}, 159--180.

\bibitem[{Kulyk et~al.(2022)Kulyk, Rafnsson, Borberg, and Pedersen}]{ddcc67d7ea224a9ebe46ab14be1f8600}
Kulyk, O.; Rafnsson, W.; Borberg, I.; and Pedersen, R. 2022.
\newblock “So I Sold My Soul'': Effects of Dark Patterns in Cookie Notices on End-User Behavior and Perceptions.
\newblock In \emph{SOUPS}.

\bibitem[{Lorenzo(2015)}]{cab}
Lorenzo, F.-B. 2015.
\newblock Redditor cracks anonymous data trove to pinpoint Muslim cab drivers.

\bibitem[{Malhotra, Kim, and Agarwal(2004)}]{malhotra_internet_2004}
Malhotra, N.~K.; Kim, S.~S.; and Agarwal, J. 2004.
\newblock Internet {Users}' {Information} {Privacy} {Concerns} ({IUIPC}): {The} {Construct}, the {Scale}, and a {Causal} {Model}.
\newblock \emph{Information Systems Research}, 15(4): 336--355.

\bibitem[{Martin and Nissenbaum(2019)}]{martin_what_2019}
Martin, K. E.~M.; and Nissenbaum, H.~F. 2019.
\newblock What {Is} {It} {About} {Location}?
\newblock \emph{SSRN Electronic Journal}.

\bibitem[{Mikians et~al.(2012)Mikians, Gyarmati, Erramilli, and Laoutaris}]{10.1145/2390231.2390245}
Mikians, J.; Gyarmati, L.; Erramilli, V.; and Laoutaris, N. 2012.
\newblock Detecting Price and Search Discrimination on the Internet.
\newblock In \emph{Workshop on Hot Topics in Networks}, 79–84.

\bibitem[{Mink et~al.(2022)Mink, Yuile, Pal, Aviv, and Bates}]{10.1145/3491102.3502136}
Mink, J.; Yuile, A.~R.; Pal, U.; Aviv, A.~J.; and Bates, A. 2022.
\newblock Users Can Deduce Sensitive Locations Protected by Privacy Zones on Fitness Tracking Apps.
\newblock In \emph{CHI}.

\bibitem[{Naeini et~al.(2017)Naeini, Bhagavatula, Habib, Degeling, Bauer, Cranor, and Sadeh}]{naeini_privacy_2017}
Naeini, P.~E.; Bhagavatula, S.; Habib, H.; Degeling, M.; Bauer, L.; Cranor, L.; and Sadeh, N. 2017.
\newblock Privacy {Expectations} and {Preferences} in an {IoT} {World}.
\newblock In \emph{SOUPS}.

\bibitem[{Poikela and Kaiser(2016)}]{EuroUSEC16}
Poikela, M.~E.; and Kaiser, F. 2016.
\newblock ”It Is a Topic That Confuses Me” – Privacy Perceptions in Usage of Location-Based Applications.
\newblock In \emph{EuroUSEC}.

\bibitem[{Primault et~al.(2019)Primault, Boutet, Mokhtar, and Brunie}]{8482357}
Primault, V.; Boutet, A.; Mokhtar, S.~B.; and Brunie, L. 2019.
\newblock The Long Road to Computational Location Privacy: A Survey.
\newblock \emph{IEEE Communications Surveys \& Tutorials}, 21(3): 2772--2793.

\bibitem[{Rebekah and Rachel(2016)}]{journals/popets/Overdorf16}
Rebekah, O.; and Rachel, G. 2016.
\newblock Blogs, Twitter Feeds, and Reddit Comments: Cross-domain Authorship Attribution.
\newblock \emph{PoPETS}, 2016(3): 155--171.

\bibitem[{Riederer et~al.(2016{\natexlab{a}})Riederer, Kim, Chaintreau, Korula, and Lattanzi}]{10.1145/2872427.2883002}
Riederer, C.; Kim, Y.; Chaintreau, A.; Korula, N.; and Lattanzi, S. 2016{\natexlab{a}}.
\newblock Linking Users Across Domains with Location Data: Theory and Validation.
\newblock In \emph{WWW}, 707–719.

\bibitem[{Riederer et~al.(2016{\natexlab{b}})Riederer, Echickson, Huang, and Chaintreau}]{Riederer2016FindYouAP}
Riederer, C.~J.; Echickson, D.; Huang, S.; and Chaintreau, A. 2016{\natexlab{b}}.
\newblock FindYou: A Personal Location Privacy Auditing Tool.
\newblock \emph{WWW}.

\bibitem[{Sadilek and Krumm(2012)}]{sadilek2012far}
Sadilek, A.; and Krumm, J. 2012.
\newblock Far Out: Predicting Long-Term Human Mobility.
\newblock In \emph{AAAI}.

\bibitem[{Sharad and Danezis(2014)}]{10.1145/2665943.2665960}
Sharad, K.; and Danezis, G. 2014.
\newblock An Automated Social Graph De-Anonymization Technique.
\newblock In \emph{WPES}, 47–58.

\bibitem[{Stevens and D’Hondt(2010)}]{stevens2010crowdsourcing}
Stevens, M.; and D’Hondt, E. 2010.
\newblock Crowdsourcing of pollution data using smartphones.
\newblock In \emph{Workshop on ubiquitous crowdsourcing}, 1--4.

\bibitem[{Stuart~A. and Charlie(2019)}]{trump}
Stuart~A., T.; and Charlie, W. 2019.
\newblock How to Track President Trump.

\bibitem[{Velykoivanenko et~al.(2021)Velykoivanenko, Niksirat, Zufferey, Humbert, Huguenin, and Cherubini}]{DBLP:journals/imwut/VelykoivanenkoN21}
Velykoivanenko, L.; Niksirat, K.~S.; Zufferey, N.; Humbert, M.; Huguenin, K.; and Cherubini, M. 2021.
\newblock Are Those Steps Worth Your Privacy?: Fitness-Tracker Users' Perceptions of Privacy and Utility.
\newblock \emph{IMWUT}, 5(4): 181:1--181:41.

\bibitem[{Veys et~al.(2021)Veys, Serrano, Stamos, Herman, Reitinger, Mazurek, and Ur}]{274435}
Veys, S.; Serrano, D.; Stamos, M.; Herman, M.; Reitinger, N.; Mazurek, M.~L.; and Ur, B. 2021.
\newblock Pursuing Usable and Useful Data Downloads Under {GDPR/CCPA} Access Rights via {Co-Design}.
\newblock In \emph{SOUPS}, 217--242.

\bibitem[{Wijesekera et~al.(2015)Wijesekera, Baokar, Hosseini, Egelman, Wagner, and Beznosov}]{190982}
Wijesekera, P.; Baokar, A.; Hosseini, A.; Egelman, S.; Wagner, D.; and Beznosov, K. 2015.
\newblock Android Permissions Remystified: A Field Study on Contextual Integrity.
\newblock In \emph{USENIX Security}, 499--514.

\bibitem[{Zang and Bolot(2011)}]{Zang:2011:ALD:2030613.2030630}
Zang, H.; and Bolot, J. 2011.
\newblock Anonymization of Location Data Does Not Work: A Large-scale Measurement Study.
\newblock In \emph{MobiCom}, 145--156.

\bibitem[{Zhang and Wang(2016)}]{zhang2016inferring}
Zhang, S.; and Wang, Z. 2016.
\newblock Inferring passenger denial behavior of taxi drivers from large-scale taxi traces.
\newblock \emph{PloS one}, 11(11): e0165597.

\bibitem[{Zhong et~al.(2015)Zhong, Yuan, Zhong, Zhang, and Xie}]{10.1145/2684822.2685287}
Zhong, Y.; Yuan, N.~J.; Zhong, W.; Zhang, F.; and Xie, X. 2015.
\newblock You Are Where You Go: Inferring Demographic Attributes from Location Check-Ins.
\newblock In \emph{WSDM}, 295–304.

\bibitem[{Zufferey et~al.(2023)Zufferey, Niksirat, Humbert, and Huguenin}]{DBLP:journals/popets/ZuffereyNHH23}
Zufferey, N.; Niksirat, K.~S.; Humbert, M.; and Huguenin, K. 2023.
\newblock "Revoked just now!" Users' Behaviors Toward Fitness-Data Sharing with Third-Party Applications.
\newblock \emph{PETS}, 2023(1): 47--67.

\end{thebibliography}

\appendix

\section{Questionnaires' content}
\label{app:questionnaires}
This section provides an exhaustive presentation of the two questionnaires mentioned in the present research paper.

\subsection{Behavior questionnaire}
\label{app:study}
\subsubsection*{Sec. A: Introduction}
% ~\newline
\paragraph{Description of the study}
You are invited to participate in a research survey about location data sharing. 
If you participate, your answers will be kept confidential. 
Also, we do not collect personally identifying information such as your name and e-mail address. 
All data will be stored on a secured server and only researchers participating in this study will have access to it. 
The results of this research study might be published in scientific journals or conferences. Any published information will be aggregated and/or anonymized.

\paragraph{Consent form}
If you wish to participate in this research study, please select the “Agree” option to continue. 
It will indicate that you will answer all questions truthfully, and that you consent that we use the collected data for research purpose. 
If you select “Disagree” you will not participate in this research survey.

\subsubsection*{Sec. B: Data Sharing}
\begin{itemize}
    \item[Q1] Have you ever granted access to your location data to any mobile app? \textit{Yes/no}
    \item[Q2] What type of smartphone are you using?
    \begin{itemize}
        \item iOS
        \item Android
    \end{itemize}
    \item[Q3] Off the top of your head (i.e., without checking on your smartphone), how many mobile apps currently have access to your location data?
    \begin{itemize}
        \item 0
        \item between 1 and 5
        \item between 5 and 10
        \item more than 10
    \end{itemize}
    \item[Q4] Off the top of your head (i.e., without checking on your smartphone), please write the mobile apps that currently have access to your location data (one app per line). \textit{Free text}
    \item[Q5] Do you know how to see which apps are tracking your location? \textit{Yes/no}
    \item[Q6] Please open the setting of your phone and check if you allow location tracking.
    \begin{itemize}
        \item Enabled
        \item Disabled
    \end{itemize}
    \item[Q7] Please open the setting of your phone and check if the location enhancement via Wi-Fi and Bluetooth is enabled.
    \begin{itemize}
        \item Enabled through Wi-Fi
        \item Enabled through Bluetooth
        \item Disabled
    \end{itemize}
    \item[Q8] Please open the setting of your phone and check how many mobile apps currently have access to your location data.
    \begin{itemize}
        \item Always \textit{Number required}learning
        \item When using the app \textit{Number required}
    \end{itemize}
    \item[Q9] After checking your setting, please list the names of the mobile apps that currently have access to your location data (one app per line).
    \begin{itemize}
        \item Always \textit{Free text}
        \item When using the app \textit{Free text}
    \end{itemize}
    \item[Q10] Are you still actively using all these apps?
    \begin{itemize}
        \item Yes, I actively use all of these apps.
        \item I actively use most of these apps.
        \item I actively use only some of these apps.
        \item No, I actively use none of these apps.
    \end{itemize}
    \item[Q11] If you mentioned that you are not actively using one or several mobile apps that currently have access to your location data. Please explain why you did not revoke their access. \textit{Free text}
\end{itemize}

%\noindent\lipsum[3][1-3]

\subsubsection*{Sec. C: Inference}
\begin{itemize}
    \item[Q12] How do you think your location may be collected or inferred? \textit{Multiple choices}
    \begin{itemize}
        \item Through GPS
        \item Though Bluetooth
        \item Through Wi-Fi
        \item Through motion sensors
        \item Through the microphone
        \item Through the light sensor
        \item Through your Internet access
    \end{itemize}
    \item[Q13] Do you think it is possible to infer the following information from the location?
    \begin{itemize}
        \item Working place(s)
        \item Home place 
        \item Genomics data
        \item Religious belief	
        \item Political affinity
        \item Personality traits	
        \item Gender
        \item Mood	
        \item Biometric data
        \item Sexual orientation	
        \item Health condition
        \item Political Views	
        \item Syndical membership
    \end{itemize}
    \item[Q14] Why do you think your location might be of interest to a service provider? \textit{Multiple choices}
    \begin{itemize}
        \item To provide a service
        \item To sell them to marketers
        \item To personalize your experience
        \item Other \textit{Free text}
    \end{itemize}
    \item[Q15] Do you think you can be re-identified through your location data? \textit{Yes/no}
    \item[Q16] If technologies would reduce and control the personal information disclosed through location data, do you use them? \textit{Yes/no}
    \item[Q17] Did you check if such technologies exist and are available? \textit{Yes/no}
\end{itemize}

\subsubsection*{Sec. D: Authorisation and Control}
\begin{itemize}
    \item[Q18] Do you think it is possible to revoke (cancel) previously granted mobile app access?  \textit{Yes/no}
%    \begin{itemize}
%        \item Yes
%        \item No
%    \end{itemize}
    \item[Q19] Have you ever modified the privacy settings to change the availability of your location data?
    \begin{itemize}
        \item Yes – I increased the availability of some of the data
        \item Yes – I decreased the availability of some of the data
        \item No
    \end{itemize}
    \item[Q20] How often have you revoked mobile app access to your location data?
    \begin{itemize}
        \item Never
        \item Only once
        \item 2-5 times
        \item 6-10 times
        \item More than 10 times
    \end{itemize}
    \item[Q21] If you already revoked access to those apps, why did you do it? \textit{Free text}
    \item[Q22] Imagine that you granted access to a mobile app and you agreed to share all the data that it is possible to share. Select which of the following data you think the third party app has access to. \textit{Multiple choices}
    \vspace{-3.5mm}
    \begin{itemize}
        \item User location
        \item User’s contacts
        \item SMS messages
        \item MMS messages
        \item User’s phone number
        \item Email address
        \item Username
        \item Address
        \item Camera
        \item Audio
        \item Phone call logs
        \item Storage
        \item User’s contacts 
    \end{itemize}
     %   \begin{itemize}
      %  \item User location \tabto{3.5cm}  --  User’s contacts
       % \item SMS messages \tabto{3.5cm}  --  MMS messages
        %\item User’s phone number \tabto{3.5cm}  -- Email address
        %\item Username \tabto{3.5cm}  --  Address
        %\item Camera \tabto{3.5cm}  --  Audio
        %\item Phone call logs \tabto{3.5cm}  --  Storage
        %\item User’s contacts 
    %\end{itemize}
    \item[Q23] Usually, when a mobile application ask you to grant access your location, what do you select?
    \begin{itemize}
        \item I refuse
        \item I accept only this time
        \item When using the app
    \end{itemize}
    \item[Q24] How difficult do you find it to monitor or revoke access granted to mobile apps?
    \begin{itemize}
        \item Very easy
        \item Easy
        \item Moderately easy
        \item Neutral
        \item Moderately difficult
        \item Difficult
        \item Very difficult
    \end{itemize}
    \item[Q25] What are your suggestions to facilitate the process of monitoring, granting, or revoking the access to mobile apps? \textit{Free text}
\end{itemize}

\subsubsection*{Sec. E: IUIPC}
Internet User’s Information Privacy Concerns based on a 7 Likert scale, see \cite{malhotra_internet_2004}.

\subsubsection*{Sec. F: Demographics}
\begin{itemize}
    \item[Q26] With which gender do you identify the most?
    \begin{itemize}
        \item Woman
        \item Man
        \item Non-binary
        \item Prefer to self describe
        \item Prefer not to answer 
    \end{itemize}
    \item[Q27] How old are you? \textit{Free text}
\end{itemize}

% \subsubsection{Sec. G: Mental Model}
% An optional section of the questionnaire asked the participants ``to draw a picture representing how you think the access granting to mobile apps is processed, and how your location data is transferred between different entities.''.
% This part was however almost not answered at all, and therefore omitted in the presentation of our results.
%\vspace*{2cm}
\subsection{Perception questionnaire}
\label{app:perception}
\begin{itemize}
    \item[Q28] Have you analyzed your own location traces?\footnote{This question was asked to ensure the validity of their traces, but note that participants had nonetheless access to artificially generated traces to explore the demonstration platform.} \textit{Yes/no}
    \item[Q29] Are you surprised by the number of inferences possible through location data? \textit{Yes/no}
    \item[Q30] Do you think Google is analyzing your location data? \textit{Yes/no}
    \item[Q31] Do you think Google shares your location data with third parties? \textit{Yes/no}
    \item[Q32] How do you think location data could be misused by a malicious person? \textit{Free text}
    \item[Q33] Can you cite any scandals or cases reported by the press in relation to these risks? \textit{Free text}
    \item[Q34] Do you think you can be re-identified through your location data? \textit{Yes/no}
    \item[Q35] Do you think you are well protected against these risks? \textit{Yes/no}
    \item[Q36] What are the measures imposed by the GDPR that reduce these risks? \textit{Free text}
    \item[Q37] Have you planed to change your privacy settings?
    \begin{itemize}
        \item Yes \textit{If yes, free text}
        \item No
    \end{itemize}
    \item[Q38] Have you planed to better use privacy-preserving technologies? \textit{Yes/no}
    \item[Q39] What privacy-preserving tools do you know? \textit{Free text}
    \item[Q40] Would you be interested in more control over the information learned for each application?
    \begin{itemize}
        \item Yes \textit{If yes, free text}
        \item No
    \end{itemize}
\end{itemize}

\newpage
\section{Codebooks}
\label{app:codebooks}

This section provides codebooks for the thematic analysis on the free text answered in the study.

% \usepackage{multirow}
\begin{table}[hbt!]
\begin{tabular}{|l|l|}
\hline
\multicolumn{1}{|c|}{\textbf{Themes}} & \multicolumn{1}{c|}{\textbf{Codes}} \\ \hline
\multirow{2}{*}{\textit{Physical harms}} & Physical harm \\ \cline{2-2} 
 & Harassment \\ \hline
\multirow{4}{*}{\textit{Economic harms}} & Blackmail \\ \cline{2-2} 
 & Robbery \\ \cline{2-2} 
 & Scam \\ \cline{2-2} 
 & Identity theft \\ \hline
\textit{Reputational harms} & Doxing \\ \hline
\multicolumn{1}{|c|}{\multirow{5}{*}{\textit{Informational integrity}}} & Surveillance \\ \cline{2-2} 
\multicolumn{1}{|c|}{} & Stalking \\ \cline{2-2} 
\multicolumn{1}{|c|}{} & Inference \\ \cline{2-2} 
\multicolumn{1}{|c|}{} & Data selling \\ \cline{2-2} 
\multicolumn{1}{|c|}{} & Targeted advertising \\ \hline
\textit{Psychological harms} & Manipulation \\ \hline
\end{tabular}
\vspace{3mm}
\caption{Codebook for the thematic analysis performed on the answers to \textit{Q32} ``How do you think location data could be misused by a malicious person?'' }
\label{tab:harms}
%\vspace{-5.5mm}
\end{table}

% \usepackage{multirow}
\begin{table}[hbt!]
\begin{tabular}{|l|l|}
\hline
\multicolumn{1}{|c|}{\textbf{Categories}} & \multicolumn{1}{c|}{\textbf{Codes}} \\ \hline
\multirow{4}{*}{\textit{Rights}} & Right to be forgotten \\ \cline{2-2} 
 & Right to access \\ \cline{2-2} 
 & Right to rectify \\ \cline{2-2} 
 & Right to object \\ \hline
\textit{Consent} &  \\ \hline
\textit{Clear purpose and transparency} &  \\ \hline
\multirow{3}{*}{\textit{Miscellaneous}} & Data security \\ \cline{2-2} 
 & Information \\ \cline{2-2} 
 & Limit on retention time \\ \hline
\end{tabular}
\vspace{3mm}
\caption{Codebook for the categorization performed on the answers to \textit{Q36} ``What are the measures imposed by the GDPR that reduce these risks?''}
\label{tab:gdpr}
\end{table}

\end{document}